\documentclass[journal]{IEEEtran}

\usepackage{color,verbatim}
\usepackage{graphicx}
\usepackage{amsfonts}
\usepackage{amsmath}
\usepackage{mathtools}
\usepackage{amssymb}
\usepackage{bm}
\usepackage{color,verbatim}
\usepackage{multirow}
\usepackage{accents}
\usepackage{theoremref}
\usepackage{tabularx}
\usepackage{algorithmic}
\usepackage{algorithm}
\usepackage{url}
\usepackage{array}
\usepackage[normalem]{ulem}
\usepackage{amsthm}
\usepackage[table]{xcolor}
\usepackage{diagbox}
\usepackage{caption}
\usepackage{subcaption}
\usepackage[hidelinks]{hyperref}

\definecolor{orange}{rgb}{1,0.8,0}
\definecolor{gray}{rgb}{.9,0.9,0.9}
\definecolor{darkgray}{rgb}{.3,0.3,0.3}
\definecolor{darkblue}{rgb}{.1,0.0,0.3}
\definecolor{lightblue}{rgb}{0.7,0.7,1}
\definecolor{lightred}{rgb}{1,0.7,.7}
\definecolor{purple}{RGB}{204,153,255}
\definecolor{lightgray}{rgb}{.95,0.95,0.95}
\definecolor{lightgreen}{rgb}{0.3,0.5,0.3}
\definecolor{darkgreen}{rgb}{0.05,0.3,0.05}

\newcommand*{\greysquare}{\textcolor{gray}{\blacksquare}}
\newcommand{\hslab}{\tilde{\mathcal{H}}^{(n)}_t}
\newcommand{\hplus}{\hat{H}^{(n)}_t}
\newcommand{\hmin}{\check{H}^{(n)}_t}

\newtheoremstyle{paperstyle} 
  {5pt} 
  {5pt} 
  {} 
  {} 
  {\bfseries} 
  {.} 
  { } 
  {\thmname{#1}\thmnumber{ #2}\thmnote{ (#3)}} 

\theoremstyle{paperstyle}

\begin{document}

    \title{Consistent Signal Reconstruction from Streaming Multivariate Time Series\thanks{This work was supported by the SFI Offshore Mechatronics grant 237896/O30, the IKTPLUSS DISCO grant 338740 from the Norwegian Science Foundation and the VALIDATE project grant 101057263 from the EU HORIZON-RIA.}}
    \author{
    \IEEEauthorblockN{Emilio Ruiz-Moreno\IEEEauthorrefmark{1}\IEEEauthorrefmark{2}, Luis Miguel L\'opez-Ramos\IEEEauthorrefmark{1}\IEEEauthorrefmark{3},~\IEEEmembership{Member,~IEEE},\\and Baltasar Beferull-Lozano\IEEEauthorrefmark{1}\IEEEauthorrefmark{2},~\IEEEmembership{Senior Member,~IEEE}\\
    \vspace{5pt}
    \IEEEauthorblockA{\IEEEauthorrefmark{1}WISENET Center, Department of ICT, University of Agder, Grimstad, Norway}\\
    \IEEEauthorblockA{\IEEEauthorrefmark{2} SIGIPRO Department, Simula Metropolitan Center for Digital Engineering, Oslo, Norway} \\
    \IEEEauthorblockA{\IEEEauthorrefmark{3} HOST Department, Simula Metropolitan Center for Digital Engineering, Oslo, Norway}
    }
    \vspace{-10pt}
    }
    \maketitle
    
\begin{abstract}
Digitalizing real-world analog signals typically involves sampling in time and discretizing in amplitude.
Subsequent signal reconstructions inevitably incur an error that depends on the amplitude resolution and the temporal density of the acquired samples.
From an implementation viewpoint, consistent signal reconstruction methods have proven a profitable error-rate decay as the sampling rate increases.
Despite that, these results are obtained under offline settings.
Therefore, a research gap exists regarding methods for consistent signal reconstruction from data streams.
Solving this problem is of great importance because such methods could run at a lower computational cost than the existing offline ones or be used under real-time requirements without losing the benefits of ensuring consistency.
In this paper, we formalize for the first time the concept of consistent signal reconstruction from streaming time-series data.
Then, we present a signal reconstruction method able to enforce consistency and also exploit the spatiotemporal dependencies of streaming multivariate time-series data to further reduce the signal reconstruction error.
Our experiments show that our proposed method achieves a favorable error-rate decay with the sampling rate compared to a similar but non-consistent reconstruction.
\end{abstract}

\section{Introduction} \label{sec:introduction}
\begin{figure}
\begin{subfigure}{\columnwidth}
    \centering
    \includegraphics[width=0.98\linewidth]{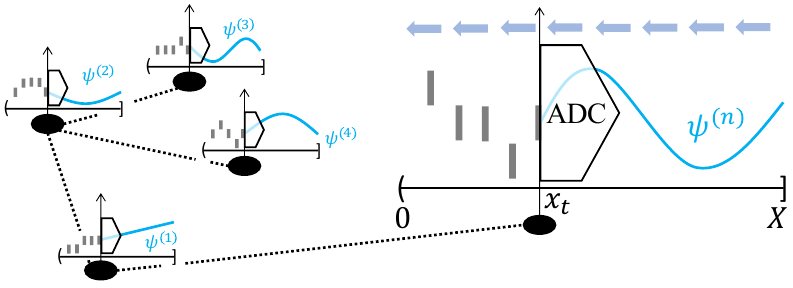}
    \caption{Snapshot at time $x_t$ of a multivariate signal being converted from analog to digital. The acquired data samples form a streamed multivariate time series of quantization intervals.}
    \label{subfig:acquisition}
\end{subfigure}
\hfill
\begin{subfigure}{\columnwidth}
    \centering
    \includegraphics[width=0.98\linewidth]{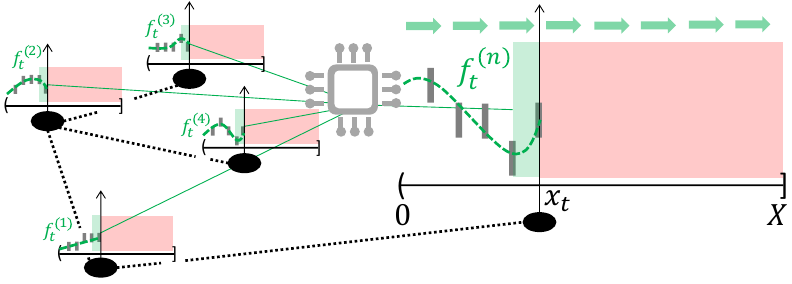}
    \caption{Snapshot at time $x_t$ of the multivariate time series from Fig. \ref{subfig:acquisition} being reconstructed. The consistent signal reconstruction method is represented visually as a microprocessor. The green shaded area indicates the currently reconstructed portion of the signal, while the red represents the future.}
    \label{subfig:reconstruction}
\end{subfigure}
\caption{Illustration of the considered signal acquisition-reconstruction process spanning from time $0$ to $X$. Stationary spatial relationships are represented as edges in a graph with black dotted lines.}
\label{fig:intro}
\end{figure}

Most real-world physical quantities can only be observed by sampling their continuous-time waveforms (analog signals). 
If the analog signal is bandlimited, i.e., its spectral density has bounded support, we can ensure a reversible discretization in time by sampling at, or above, its Nyquist sampling rate \cite{jerri1977shannon,luke1999origins}. 
This is a universal sufficient condition but not necessary for all signals: some non-bandlimited signals can be perfectly reconstructed after uniformly sampled at, or above, its finite rate of innovation.
Examples of such signals are streams of Dirac delta functions and \emph{splines} \cite{vetterli2002sampling}.
Nonetheless, we can only observe a modified version of the sampled analog signal in practice because of the physical limitations of acquisition systems.
For instance, acquisition systems such as analog-to-digital converters (ADCs) introduce both a sampling and a quantization step.
The consequent discretization in amplitude causes an irreversible loss of information, making perfect signal reconstruction no longer possible.

Even though quantization is a deterministic operation, the signal reconstruction error-rate decay dependencies can be analyzed effectively from a stochastic point of view.
For example, the reconstruction error caused by a uniform quantizer can be well approximated and modeled as an additive uniform white noise source independent of the input signal when certain assumptions are met \cite{bennett1948spectra,gray1990quantization}.
Under these assumptions, the quantization noise spectral density is constant. 
Thus, if an analog signal $\psi$ is bandlimited, we can reduce the amount of in-band quantization noise by sampling above its Nyquist sampling rate (oversampling).
Accordingly, we can reduce the reconstruction error by digitally filtering out the quantization noise above the original signal bandwidth.
In this scenario, it is well known that low-pass filtering reduces the mean squared error (MSE) of a reconstructed signal $f$ by a factor proportional to the squared quantization step size $\Delta^2$ and inversely proportional to the oversampling ratio $R$.
Formally,
\begin{equation} \label{eq:white_noise_decay}
    \mathbb{E} \left[ \left\Vert f - \psi \right\Vert^2_2 \right] \propto \frac{\Delta^2}{R} ,
\end{equation}
where the expectation is taken over the amplitude of the reconstruction error, modeled as a uniform random variable \cite{bennett1948spectra}.
The result shown in (\ref{eq:white_noise_decay}) is useful for identifying the main factors that influence the signal reconstruction error-rate decay under an oversampling regime. 
However, it does not reflect the effects of any quantization scheme or reconstruction method other than uniform quantization and linear decoding, i.e., low-pass filtering of the quantized signal at a cut-off frequency.

In contrast, deterministic analyses, based on deterministic bounds over the quantization levels, show that it is possible to reconstruct quantized signals with a squared norm error that is asymptotically proportional to the square of the quantization step size and inversely proportional to the square of the oversampling ratio, that is
\begin{equation} \label{eq:deterministic_error-rate_decay}
    \left\Vert f - \psi \right\Vert^2_2 = \mathcal{O} \left( \frac{\Delta^2}{R^2} \right).
\end{equation}
The trend in (\ref{eq:deterministic_error-rate_decay}) can be achieved by performing signal reconstruction via \emph{consistent signal estimates}, where the term ``consistent'' here means coherent with all available knowledge about the original analog signal and the acquisition system.
This has been shown for bandlimited signals \cite{thao1994deterministic,thao1994reduction}, and non-bandlimited signals with finite rate of innovation \cite{jovanovic2006oversampled}.
The behavior presented in (\ref{eq:deterministic_error-rate_decay}) suggests that when a uniformly quantized and oversampled signal is reconstructed using consistent estimates, the reconstruction error can be asymptotically reduced at the same rate by either increasing the oversampling ratio or decreasing the quantization step size.
This balanced signal reconstruction error-rate decay has important implications.
For instance, in practice, it is more convenient to increase the oversampling ratio because the implementation cost of the analog circuitry necessary for halving the quantization step size is much higher than that for doubling the oversampling ratio \cite{nguyen1993deterministic}.

Notwithstanding the above, consistent signal reconstructions usually require sophisticated methods that are far more computationally complex than linear decoding, thus hindering their practical implementation.
Only a few previous studies have explored how to alleviate the computational complexity of yielding consistent reconstructions while maintaining the asymptotic behavior established in (\ref{eq:deterministic_error-rate_decay}).
For example, in \cite{beferull2003efficient}, linear decoding computational complexity is achieved by using different quantization step sizes, while the difficulty lies in the design of the quantization scheme.
As another example, the authors of \cite{rangan2001recursive} design a subtractive dithering method aimed at recursively solving an overdetermined system of linear equations set up from quantized data described by frames\footnote{Frames provide redundant and usually non-unique representations of vectors \cite{goyal2001quantized,kovacevic2013fourier}.}.
However, the resulting signal estimates are only guaranteed to be consistent with the most recently analyzed frame representation in the recursion. 
In summary, there seems to be a research gap regarding low-complexity consistent signal reconstruction methods under uniform quantization schemes.

Online schemes are a reasonable approach to fill this research gap as they allow for a low runtime complexity per data sample.
They can also be used to deal with streaming data sequentially, possibly under real-time requirements, in exchange for sacrificing some accuracy in the solution.
To the best of our knowledge, there is an absence of methodological frameworks dedicated to consistent signal reconstruction methods designed under online settings or for sequentially reconstructing streaming data samples.
Consequently, there are no studies about their error-rate decay dependencies either.

This paper aims to cover the aforementioned gap.
To this end, we focus on a class of multivariate, non-bandlimited, temporally and spatially interdependent smooth signals with a finite rate of innovation that can be expressed in terms of splines.
This is justified because most signals of interest are smooth, usually elapse within a finite timespan, and rarely occur in complete isolation.
Then, we consider an ADC that uniformly samples and quantizes these multivariate signals, resulting in streamed multivariate time series of quantization intervals, as illustrated in Fig. \ref{subfig:acquisition}.
From here, our goal is to design a consistent signal reconstruction approach from streaming data that satisfies two key requirements: i) signal smoothness, in agreement with the signals studied in this work, and ii) a zero-delay response, i.e., the ability to process each streaming data sample as it arrives and before the next sample is available.
The considered requirements are of great importance in practice since they arise naturally in a wide range of signal reconstruction problems, ranging from high-speed digital-to-analog conversion \cite{schmidt2017high} to online trajectory planning \cite{bazaz1999minimum,marcucci2023fast}, among others.

Concretely, this paper presents the first approach to zero-delay consistent signal reconstruction from a streaming multivariate time series of quantization intervals (see Fig. \ref{subfig:reconstruction}), as well as an experimental study of its error-rate decay dependencies.
Specifically, our approach follows a similar strategy as the one presented in \cite{ruizmoreno2023trainable}, which describes how to formulate a smoothing spline interpolation problem from streaming data under a sequential decision-making perspective and zero-delay requirements.
In \cite{ruizmoreno2023trainable}, a stationary parametric policy, i.e., a time-invariant decision strategy, based on tunable parameters on top of a recurrent neural network (RNN) \cite{salehinejad2017recent} is devised.
Then, the policy is trained over a set of representative streaming data to reduce a given trade-off between the sum of squared residuals and the roughness, i.e., a derivative-based measure of the model complexity \cite{craven1978smoothing}, of the reconstructed signal.
However, that strategy is intended for univariate and noisy data samples and fails to ensure consistency in the reconstruction as well as to exploit the spatial dependencies of multivariate streaming data.

The work in this paper considerably differs from the work in \cite{ruizmoreno2023trainable}.
This is because our zero-delay consistent signal reconstructions are obtained by optimizing a different objective than the one introduced in \cite{ruizmoreno2023trainable}.
Furthermore, the resulting optimization problem contains additional constraints to ensure consistency, such as enforcing the reconstructed signal to pass through quantization intervals, thus requiring seeking solutions in different feasible sets.
These facts substantially change the procedure in which the solution is found.
Moreover, the multivariate analysis presented in this paper allows us to reach additional results and conclusions that would not be available after merely extending the approach in \cite{ruizmoreno2023trainable} to multivariate signals.
Concretely, offline consistent signal reconstructions are not challenged by the uncertainty of the to-be-received data samples and hence do not benefit from a multivariate formulation (due to separability), whereas our zero-delay consistent signal reconstruction does improve (in terms of the error-rate decay) when incorporating the spatial relations among the multiple time series of quantization intervals.
Therefore, the extension to the multivariate case provides new insights that improve our understanding of online consistent signal reconstructions.

The main contributions of this paper can be summarized as follows:
\begin{itemize}
    \item We formalize the concept of consistent signal reconstruction from streaming data under zero-delay response requirements.
    As a result, our formulation generalizes the concept of consistency beyond offline settings.
    \item To the best of our knowledge, this work is the first to devise a consistent method for zero-delay signal reconstruction from streaming multivariate time series.
    The method can enforce consistency in a closed-form step allowing for a faster and more convenient implementation than numerical alternatives.
    \item We show experimentally that the reconstruction error incurred by our proposed method decays at the same rate by decreasing the quantization step size or increasing the oversampling ratio. 
    Moreover, the error-rate decay slope with respect to the oversampling ratio doubles (in logarithmic scale) the one obtained with a similar zero-delay smooth signal reconstruction method that does not enforce consistency.
    \item Additionally, we observe that the error-rate decay incurred by our method does improve when incorporating the spatial relationships among the multiple time series, whereas offline consistent signal reconstructions do not benefit from a multivariate formulation.
\end{itemize}

The rest of the paper is organized as follows. Sec. \ref{sec:notation} introduces the main notation used throughout the paper.
Sec. \ref{sec:consistency} formalizes the notion of consistency and presents the concept of consistent signal reconstruction from streaming data under zero-delay requirements.
Then, Sec. \ref{sec:interpolation} describes our proposed method from a sequential decision-making perspective, and Sec. \ref{sec:empirical} analyzes its error-rate decay behavior experimentally.
Finally, Sec. \ref{sec:conclusion} concludes the paper.
    \section{Mathematical notation} \label{sec:notation}
This section introduces the mathematical notation most recurrently used throughout the paper.

Vectors and matrices are represented in bold lowercase and bold capital letters, respectively. 
Given a vector $\bm{v} = [v_1,\dots,v_C]^\top$ its \textit{c}th component is indicated as $[\bm{v}]_c \triangleq v_c$. 
Similarly, given a matrix $\bm{M}\in\mathbb{R}^{R\times C}$, the element in the \textit{r}th row and \textit{c}th column is indicated as $[\bm{M}]_{r,c}$.
The notation $[\bm{v}]_{i:j}$ refers to the sliced vector $[v_i,\dots,v_j]\in\mathbb{R}^{j-i+1}$.
We use Euler’s notation for the derivative operator; thus, $D^k_x$ denotes the \textit{k}th derivative over the variable $x$.
Lastly, we refer to functions $f$ as elements of a function space or explicitly allude to its domain and codomain with the compact notation $f(x)$, depending on the context.
\section{Consistency} \label{sec:consistency}
This section describes and models the function space comprising the signals considered in this work and the acquisition system.
Accordingly, we identify and define the set of consistent signal estimates and introduce the related concept of consistent signal reconstruction from streaming acquired data.

\subsection{Function space of multivariate smooth signals} \label{ssec:signal_space}
Most real-world physical processes are bounded and smooth due to energy conservation \cite{kosheleva2020physical}.
Moreover, most of them occur within a certain finite time lapse.
Therefore, we assume any real-world process-associated measurable quantity to be described by smooth and non-bandlimited signals.

Let us define $\mathcal{W}_\rho$ as the function space of real functions defined over the domain $(0,X]\subseteq\mathbb{R}$ with $\rho-1$ absolutely continuous derivatives and with the $\rho$th derivative square integrable.
Then, any of the aforementioned real-world signals can be accurately modeled by a function $\psi\in\mathcal{W}_\rho$.
In practice, $\psi$ can be approximated, or reconstructed, from a set of $T_*$ uniformly sampled measurements $\{\psi(tX/T_*)\}^{T_*}_{t=1}$ as follows
\begin{subequations} \label{eq:natural_spline_approximation}
\begin{align}
    f_* \triangleq \text{arg}\underset{f\in\mathcal{W}_\rho}{\text{ min }}& \int^X_{0} \left( D^\rho_x f(x) \right)^2 \, dx \label{seq:roughness}\\
    \text{subject to: }& f\left(\frac{tX}{T_*}\right) = \psi\left(\frac{tX}{T_*}\right), \,\, \forall t \in \mathbb{N}^{[1,T_*]} ,
\end{align}
\end{subequations}
where the metric used as objective in \eqref{seq:roughness} is usually referred to as roughness and $f_*$ is unique as long as $T_*\geq\rho$ \cite{schoenberg1964interpolation,green1993nonparametric}.
Actually, any function $\psi\in\mathcal{W}_\rho$ with roughness bounded by an arbitrary constant is known to satisfy
\begin{equation} \label{eq:natural_spline_approximation_upperbound}
    \underset{x\in[X/T_*,X)}{\text{ sup }} \left| D^k_x\psi(x) - D^k_x f_*(x) \right| \leq \text{const} \, \cdot \, \left(\frac{X}{T_*}\right)^{\rho-k} ,
\end{equation}
for all $k\in \{0,\dots,\rho-1\}$ \cite{wahba1990spline}.
That is, the reconstruction error, up to the $(\rho-1)$th derivative, is upper bounded by a term inversely proportional to the number $T_*$ of measurements.

Since the reconstruction error in \eqref{eq:natural_spline_approximation_upperbound} can be made arbitrarily small by increasing the number of samples $T_*$, we restrict ourselves without loss of generality to those signals in $\mathcal{W}_\rho$ whose approximation, as in \eqref{eq:natural_spline_approximation}, is exact, i.e., the left-hand term in \eqref{eq:natural_spline_approximation_upperbound} is zero-valued for all allowed \textit{k}th derivatives.
The function space where such signals of interest belong is denoted as $\mathcal{V}_\rho\subseteq\mathcal{W}_\rho$, and turns out to be constituted by splines, i.e., piecewise polynomial functions, of order $2\rho-1$ with $T_*$ knots (joint points) at $\mathcal{X}_* = \{ tX/T_* \}^{T_*}_{t=1}$ \cite{wahba1990spline}.

On the other hand, more than one signal often concurrently stems from the same physical process.
Given that, one can expect spatial interdependencies among them.
For this reason, this work deals with multivariate signals of the form $\bm{\Psi} = [\psi^{(1)},\dots,\psi^{(N)}]^\top$ with $\psi^{(n)}\in\mathcal{V}_{\rho_n}$ for all $n\in\mathbb{N}^{[1,N]}$, namely, $\bm{\Psi} \in \mathcal{V}_{\rho_1}\times\dots\times\mathcal{V}_{\rho_N}$.
However, and for the sake of simplicity in the notation, we rather focus on those multivariate signals in $\mathcal{V}_{\rho}\times\dots\times\mathcal{V}_{\rho} = \mathcal{V}^N_\rho$, i.e., same order $\rho$ across all $N$ dimensions, since the considered formulation can be easily adapted to the presented general case.

\subsection{Acquisition system and irreversibility} \label{ssec:acquisition_system}
Thanks to the reconstruction procedure presented in \eqref{eq:natural_spline_approximation}, a multivariate signal $\bm{\Psi}\in\mathcal{V}^N_\rho$ can be reversibly sampled despite being non-bandlimited.
That is, we can recover the multivariate signal $\bm{\Psi}$ from finitely many measurements as long as they contain its set of knots, namely, the signal values $\bm{\Psi}(x)$ sampled at $x\in\mathcal{X}_*$.
For example, we can measure the knots by sampling uniformly at a rate $\nu_* = T_*/X$.
In fact, any uniform sampling rate $\nu=R\nu_*$, with $R$ being a positive integer oversampling ratio, measures all the knots in $\mathcal{X}_*$ because $\mathcal{X}=\{t\nu^{-1}\}^{RT_*}_{t=1}\supseteq\mathcal{X}_*$.

Ideally, these measurements would be acquired through an infinite amplitude-resolution sampling mechanism, namely, an ideal sampler.
However, this is impossible in practice due to the physical limitations of measuring devices \cite{widrow2008quantization}.
Instead, the usual procedure is to assume an ideal sampler and then model the resulting finite amplitude-resolution measurements as the outcome of a quantization mapping.
In this work, we assume an ideal sampler working at a uniform sampling rate $\nu$ followed by a uniform midtread quantizer with a quantization step size $\Delta^{(n)}$ for each of the $n$ signals.

The resulting acquired data consists of a streaming multivariate time series of $N$ time series of length $T=RT_*$.
We refer to every \textit{t}th term of the streaming multivariate time series as an \emph{observation}, $\bm{o}_t$.
Every \textit{t}th observation is constructed from $N$ quantization intervals described by their time stamp $x_t = tX/T\in\mathcal{X}$, common for all intervals of the \textit{t}th observation, their center $y^{(n)}_t \in \mathbb{R}$, and half quantization step size $\epsilon^{(n)} = \Delta^{(n)}/2 \in \mathbb{R}_+$.
Formally, $\bm{o}_t = [x_t,\bm{y}_t^\top,\bm{\epsilon}^\top]^\top$ where $\bm{y}_t = [y^{(1)}_t,\dots,y^{(N)}_t]^\top$ and $\bm{\epsilon} = [\epsilon^{(1)},\dots,\epsilon^{(N)}]\top$.

Unfortunately, the acquisition system under consideration is irreversible.
That is, the multivariate signals in $\mathcal{V}^N_\rho$ may not be perfectly recovered from a finite set of measurements anymore if acquired as described above. 
This drawback is two-sourced.
First, the sampler works sequentially.
Thus, intermediate acquisition stages are irreversible because not all knots, i.e., the multivariate signal values at $\mathcal{X}_*$, have been measured yet.
Second, the quantization step is a many-to-one mapping, i.e., non-injective and hence, not invertible. 
Consequently, incurring a reconstruction error in this context is unavoidable.

\subsection{Consistent signal estimate} \label{ssec:consistent_signal_estimate}
\begin{figure}
\begin{subfigure}{\columnwidth}
    \centering
    \includegraphics[width=0.825\linewidth]{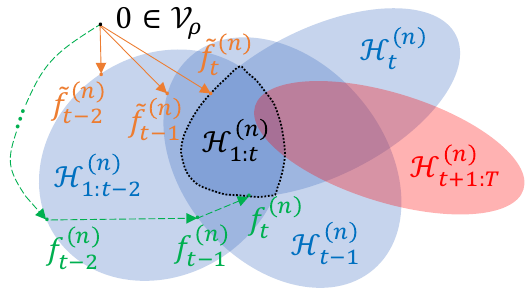}
    \caption{Intersection of hyperslabs as a Venn diagram. At time step $t$, the functions within the area enclosed by the dotted black line, $\mathcal{H}^{(n)}_{1:t}$, describe current consistent signal estimates.}
    \label{subfig:diagram}
\end{subfigure}
\hfill
\begin{subfigure}{\columnwidth}
    \centering
    \includegraphics[width=0.95\linewidth]{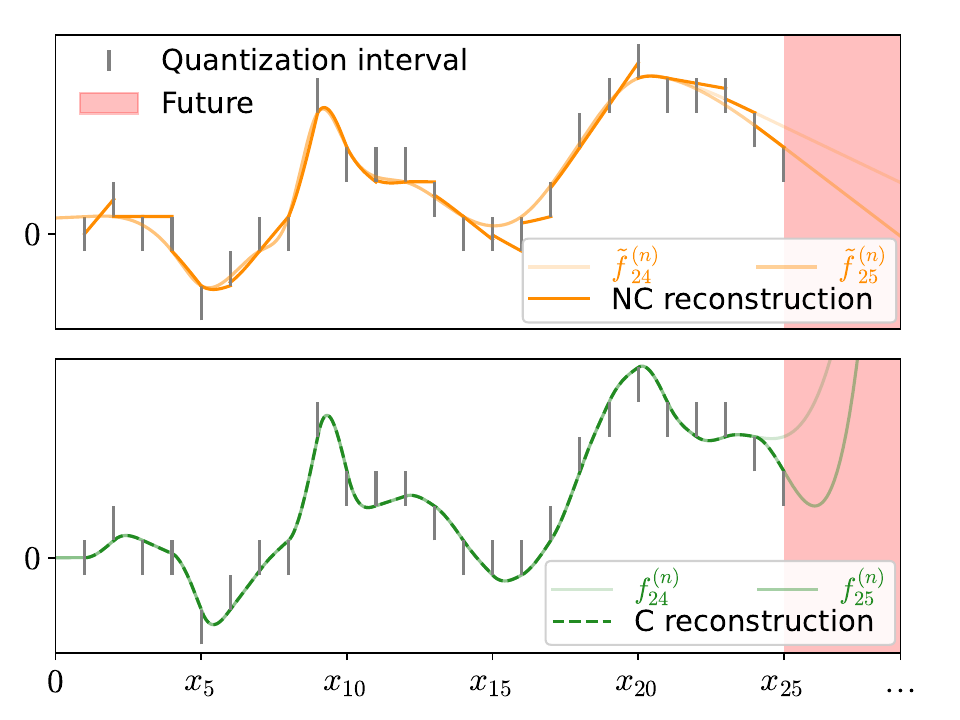}
    \caption{Comparison between the ongoing reconstructed signals using a non-consistent (NC) and a consistent (C) method. Fig. \ref{subfig:diagram} illustrates how both methods produce a sequence of consistent signal estimates, namely, $\{\Tilde{f}^{(n)}_t\}^T_{t=1}$ and $\{f^{(n)}_t\}^T_{t=1}$. However, the sequence $\{\Tilde{f}^{(n)}_t\}^T_{t=1}$ does not lead to a consistent signal reconstruction, while the sequence $\{f^{(n)}_t\}^T_{t=1}$ does.}
    \label{subfig:methods}
\end{subfigure}
\caption{Visual representation of a consistent signal estimate and a consistent signal reconstruction.}
\label{fig:consistent_signal_reconstruction}
\end{figure}
As stated in the previous Sec. \ref{ssec:acquisition_system}, it is generally impossible to recover a multivariate signal $\bm{\Psi}\in\mathcal{V}^N_\rho$ from its stream of observations.
Nevertheless, we can select a relatively reduced set of functions where the multivariate signal $\bm{\Psi}$ lies.
Specifically, for every \textit{n}th time series and for any of its \textit{i}th quantization intervals we can construct its associated hyperslab $\mathcal{H}^{(n)}_i\subseteq\mathcal{V}_\rho$ as
\begin{equation}
    \mathcal{H}^{(n)}_i = \left\{ h \in \mathcal{V}_\rho : \left|h(x_i) - y^{(n)}_i\right| \leq \epsilon^{(n)} \right\} .
\end{equation}
Then, at time step $t$, we can state that each signal $\psi^{(n)}$ belongs to the intersection of its so-far constructed hyperslabs.
That is, $\psi^{(n)}\in\mathcal{H}^{(n)}_{1:t}$ with $\mathcal{H}^{(n)}_{1:t} \triangleq \bigcap^t_{i=1}\mathcal{H}^{(n)}_i$, as illustrated in Fig. \ref{subfig:diagram}.
In this context, any signal estimate described by a function drawn from $\mathcal{H}^{(n)}_{1:t}$ is called a consistent signal estimate at time step $t$, or simply consistent signal estimate, when clear by context.
This is because any reacquisition, i.e., processing the newly estimated signal through the same acquisition system until the same time step $t$, leads to the same set of already received $t$ quantization intervals.
Equivalently, any signal estimate is conventionally referred to as consistent if and only if: i) it passes through all the (so-far) received quantization intervals, and ii) it is described by a function in $\mathcal{V}_\rho$ \cite{nguyen1993deterministic}.

Recall from Sec. \ref{ssec:signal_space} that $\mathcal{V}_\rho$ is constituted by splines of order $2\rho-1$, with $\rho-1$ continuous derivatives and $T_*$ knots located at time stamps in $\mathcal{X}_*$.
For this reason, we construct every \textit{n}th signal estimate as a spline composed of $T$ pieces, or function sections, as
\begin{equation} \label{eq:spline}
    f^{(n)}(x) = 
    \begin{cases}
        g^{(n)}_1(x), & \text{if } 0 < x \leq x_1 \\
        g^{(n)}_2(x), & \text{if } x_1 < x \leq x_2 \\
        \vdots \\
        g^{(n)}_{T}(x), & \text{if } x_{T-1} < x \leq x_{T}
    \end{cases}
\end{equation}
where every $t$th function section $g^{(n)}_t:(x_{t-1},x_t]\to\mathbb{R}$ is a linear combination of polynomials of the form
\begin{equation} \label{eq:function_section}
    g^{(n)}_t(x) = {\bm{a}^{(n)}_t}^\top \bm{p}_t(x) ,
\end{equation}
with combination coefficients $\bm{a}^{(n)}_t \in \mathbb{R}^{2\rho}$ and basis vector function $\bm{p}_t:(x_{t-1},x_t]\to\mathbb{R}^{2\rho}$ defined as
\begin{equation} \label{eq:basis_vector}
    \bm{p}_t(x) = \left[ 1,(x-x_{t-1}),\dots,(x-x_{t-1})^{2\rho-1} \right]^\top .
\end{equation}
Then, the remaining requirements for $f^{(n)}$ to belong to $\mathcal{V}_\rho$ are that the location of its knots must cover all time stamps in $\mathcal{X}_*$ and that it has to be continuous up to the $(\rho-1)$th derivative.
By design, the knots of $f^{(n)}$ lie at $\mathcal{X}$ hence, covering all time stamps in $\mathcal{X}_*$ since $\mathcal{X}_* \subseteq \mathcal{X}$.
Regarding its continuity, for $f^{(n)}$ to be $(\rho-1)$-smooth, the function sections must satisfy
\begin{equation} \label{eq:cc}
\underset{x \to x_{t-1}^-}{\text{lim }} D^k_x \, g^{(n)}_{t-1}(x) = \underset{x \to x_{t-1}^+}{\text{lim }} D^k_x \, g^{(n)}_t(x),
\end{equation}
for every $k\in [0,\rho-1]$ and $t\in[2,T]$.
This can be ensured by imposing the following equality constraint
\begin{equation} \label{eq:cc_vf}
    \left[\bm{a}^{(n)}_t\right]_{1:\rho} = \bm{e}^{(n)}_{t-1} ,
\end{equation}
for each $t\in\mathbb{N}^{[1,T]}$, where $\bm{e}^{(n)}_t \in \mathbb{R}^{\rho}$ is a vector with elements given by
\begin{equation} \label{eq:eq_constraint_element}
    [\bm{e}^{(n)}_t]_i = \frac{1}{(i-1)!}\sum^{2\rho}_{j=1}\left[\bm{a}^{(n)}_{t}\right]_j
    \, u_{t}^{j-i}
    \prod^{i-1}_{k=1}(j-k) , 
\end{equation}
where $u_t \triangleq x_t - x_{t-1}$, except $u_1=x_1$, and where $\bm{e}^{(n)}_0$ determines the initial boundary conditions of the signal estimate (\cite[Proposition 1]{ruizmoreno2023trainable}). 

In this way, any signal estimate $f^{(n)}$ constructed as in \eqref{eq:spline}, and satisfying \eqref{eq:cc}, belongs to $\mathcal{V}_\rho$.
Additionally, and as stated before, if at time step $t$ it passes through all the so-far received quantization intervals, i.e., $f^{(n)}$ satisfies $|f^{(n)}(x_i)-y^{(n)}_i|\leq\epsilon^{(n)}$ for all $i\in\{1,\dots,t\}$, or equivalently $f^{(n)}\in\mathcal{H}^{(n)}_{1:t}$, then it is a consistent signal estimate at time step $t$.

\subsection{Zero-delay consistent signal reconstruction} \label{ssec:zero-delay_consistent_signal_reconstruction}
Recall that the acquisition system modeled in Sec. \ref{ssec:acquisition_system} generates a stream of observations.
In this paper, we pursue its converse reconstruction approach; that is, our goal is to sequentially reconstruct a signal $\bm{\Psi}\in\mathcal{V}^N_\rho$ as its observations are received.
To this end, we devise a zero-delay multivariate signal reconstruction method.
In this context, our objective is to yield a sequence of signal estimates in $\mathcal{V}^N_\rho$, where the \textit{t}th multivariate signal estimate $\bm{f}_t=[f^{(1)}_t,\dots,f^{(N)}_t]^\top$ is used to reconstruct the multivariate time series between the last two received observations, i.e., $\bm{o}_{t-1}$ and $\bm{o}_t$.
Moreover, every reconstruction step has to be performed as soon as the \textit{t}th observation is received and without knowing the to-be-received observations\footnote{For the sake of completeness, implementing a zero-delay signal reconstruction method also requires a reduced constant complexity per iteration to guarantee the required execution speed and scalability \cite{wilf2002algorithms}.}.

From the discussion in Sec. \ref{ssec:consistent_signal_estimate}, one may expect to obtain a consistent signal reconstruction using a zero-delay signal reconstruction method that generates a sequence of consistent signal estimates.
That is, for each $n$th time series every $t$th signal estimate $f^{(n)}_t$ belongs to $\mathcal{H}^{(n)}_{1:t}$.
However, it is important to notice that a method producing a sequence of consistent signal estimates does not necessarily achieve a consistent reconstruction.
This is because guaranteeing a consistent reconstruction requires the sequence of signal estimates to fulfill additional requisites.
Specifically, every two consecutive consistent signal estimates, namely $f^{(n)}_{t-1}$ and $f^{(n)}_t$, must ensure smoothness at the takeover time stamp $x_{t-1}$, as shown in Fig. \ref{subfig:methods}.
Formally, a consistent signal reconstruction is attainable from a sequence of consistent signal estimates $\{f^{(n)}_t\}^T_{t=1}$ if they satisfy that $c^{(n)}_t\in\mathcal{H}^{(n)}_{1:t}$, where
\begin{equation} \label{eq:consistent_reconstruction}
    c^{(n)}_t(x) =
    \begin{cases}
        f^{(n)}_{t-1}(x) & \text{if } 0 < x \leq x_{t-1} , \\
        f^{(n)}_t(x) & \text{if } x_{t-1} < x \leq X ,
    \end{cases} 
\end{equation}
for all $t\in\mathbb{N}^{[2,T]}$. 

In our case, the presented zero-delay reconstruction methods can yield a consistent signal reconstruction by sequentially updating the spline-based signal estimate coefficients.
Concretely, if at every time step $t$, we update the remaining $T-t$ spline coefficients, i.e., $\bm{a}^{(n)}_i$ for $i=t,\dots,T$ subject to satisfying the corresponding continuity constraints in \eqref{eq:cc_vf}, and subject to passing through the last received quantization interval, i.e., the \textit{t}th spline coefficients obey $|{\bm{a}^{(n)}_t}^\top\bm{p}_t(x_t) - y^{(n)}_t|\leq\epsilon^{(n)}$, then the resulting signal estimate update $f^{(n)}_t$ is part of a sequence satisfying \eqref{eq:consistent_reconstruction}.
In this way, the overall signal reconstruction, i.e., the reconstructed signal at the last time step $T$, is completely described by the last consistent signal estimate $f^{(n)}_T$.
In practice, updating just the \textit{t}th spline coefficients at every time step $t$ (temporarily ignoring the $T-t$ remaining coefficients) eventually leads to the same consistent signal reconstruction.

Given this, notice that the core difference between the offline consistent signal reconstruction methods discussed in Sec. \ref{sec:introduction} and our approach lies in the fact that we explicitly account for the sequential nature of the acquisition system.
This is conceptually more challenging than assuming that all data samples are available in advance since in the sequential case we cannot achieve a consistent signal reconstruction with just a single consistent signal estimate, but with a sequence of them, subject to \eqref{eq:consistent_reconstruction}, instead.
This significant difference opens the question of whether a method for zero-delay consistent signal reconstruction has the corresponding benefits in the reconstruction error-rate decay behavior as the offline consistent ones have ~\cite{thao1994deterministic,thao1994reduction,beferull2003efficient,jovanovic2006oversampled}.
We anticipate a positive answer to the posed question, but before presenting such an outcome, we introduce our zero-delay consistent signal reconstruction approach in detail.

\section{Zero-delay spline interpolation} \label{sec:interpolation}
This section introduces a zero-delay spline interpolation approach based on policy training that, differently from \cite{ruizmoreno2023trainable}, can accommodate multivariate quantized data and enforce consistency as defined in Sec. \ref{sec:consistency}.

\subsection{A trainable multivariate approach to zero-delay spline interpolation} \label{ssec:multivariate_zero-delay_interpolation}
The problem of reconstructing a multivariate time series of quantization intervals using a sequence of spline-based signal estimates under smoothness and zero-delay requirements can be formalized from a sequential decision-making perspective as follows.
At time step $t$, we encode the condition of the so-far reconstructed signal, and the last received quantization interval in a vector-valued variable referred to as the \textit{t}th state $\bm{s}_t\in\mathcal{S}$, being $\mathcal{S}$ the state space.
Every \textit{t}th state is constructed as $\bm{s}_t=[x_{t-1},\bm{o}^\top_t,\bm{e}^\top_{t-1}]^\top$ with $\bm{e}_{t-1} = [{\bm{e}^{(1)}_{t-1}}^\top,\dots,{\bm{e}^{(N)}_{t-1}}^\top]^\top$ where every \textit{n}th component $\bm{e}^{(n)}_{t-1}$ represents the right hand term at $\eqref{eq:cc_vf}$.
In fact, every \textit{t}th state is fully determined once all \textit{n}th spline coefficients $\bm{a}^{(n)}_t$ are fixed, and the \textit{t}th observation $\bm{o}_t$ is received.
This can be explicitly described by a state update mechanism through a deterministic mapping, as $\bm{s}_{t+1} = F(\bm{s}_t,\bm{a}_t,\bm{o}_{t+1})$, allowing us to identify all visitable states handily.
On the other hand, selecting all \textit{n}th spline coefficients at time step \textit{t} can be understood as an action.
This is because every \textit{n}th function section $g^{(n)}_t$, as defined in \eqref{eq:function_section}, is fully determined as soon as $\bm{a}^{(n)}_t$ is chosen and hence so is the resulting piece of the subsequent signal reconstruction.
Explicitly, we construct every \textit{t}th action by concatenating all \textit{n}th spline coefficients, i.e., $\bm{a}_t = [{\bm{a}^{(1)}_t}^\top,\dots,{\bm{a}^{(N)}_t}^\top]^\top$, and denote the action space as $\mathcal{A}\subseteq\mathbb{R}^{2N\rho}$.
We distinctively indicate the set of actions accessible from a given state $\bm{s}_t$ as the admissible action set $\mathcal{A}(\bm{s}_t)\subseteq\mathcal{A}$.
In particular, we select the \textit{t}th action through a stationary parametric policy $\bm{\mu}_{\bm{\theta}}:\mathcal{S}\to\mathcal{A}$.
Stationary policies do not change over time, and they are suitable for making decisions in problems governed by stationary spatiotemporal dynamics and with varying time steps.
On the other hand, policy parametrization helps to reduce the pool of candidate policies and allows, with some flexibility, to accommodate the problem structure, e.g., spatiotemporal dynamics and constraints, into the policy.

In this context, any reconstructed signal can be assessed and therefore compared by computing a cumulative cost $K:\mathcal{S}\times\mathcal{A}\to\mathbb{R}$ through the traversed state-action pairs by following a given policy $\bm{\mu}_{\bm{\theta}}$.
From here, we can tune the policy parameters via policy training by solving the following optimization problem
\begin{subequations}  \label{eq:policy_training}
\begin{align}
    \text{arg}\underset{\bm{\theta}\in\mathbb{R}^P}{\text{ min }}& \sum_{m=1}^M \sum^{T}_{t=1} K \left( \bm{s}_{m,t},\bm{\mu}_{\bm{\theta}}(\bm{s}_{m,t}) \right) \label{seq:policy_search_objective} \\
    \text{s. to: }& \bm{s}_{m,t} = F\left(\bm{s}_{m,t-1},\bm{\mu}_{\bm{\theta}}(\bm{s}_{m,t-1}),\bm{o}_{m,t}\right), \forall m,t , \\
    & \bm{\mu}_{\bm{\theta}}(\bm{s}_{m,t}) \in \mathcal{A}(\bm{s}_{m,t}) , \forall m,t ,
\end{align}
\end{subequations}
where the integer $M$ denotes the number of example multivariate time series, indexed by \textit{m}, and where all the stream of observations $\bm{o}_{m,t}$ and initial states $\bm{s}_{m,0}$ are given.
Thus, if a sufficiently large collection of representative multivariate time series is available, one expects the trained policy to outperform, on average, any other policy with different parameter values when tested over a previously unseen stream of observations.

\subsection{Policy architectures and variations} \label{ssec:architecture_and_variation}
\begin{table}
\centering
\caption{Per time-series cost $\kappa(\cdot,\cdot)$ and admissible action set $\mathcal{A}(\cdot)$ for the consistent ($\greysquare$) and smoothing ($\square$) policy variations.}
\renewcommand{\arraystretch}{1.8}
\begin{tabular}{|c|}
\hline
Per time-series cost evaluated at the \textit{t}th state-action pair \\ \hline \hline
\cellcolor{gray} ${\bm{a}^{(n)}_t}^\top \bm{M}_t \bm{a}^{(n)}_t$ \\ \hline
$\left(\bm{p}_t^\top(x_t)\bm{a}^{(n)}_t - y^{(n)}_t\right)^2 + \eta \, {\bm{a}^{(n)}_t}^\top \bm{M}_t \bm{a}^{(n)}_t$, with $\eta>0$ \\ \hline 
\end{tabular} 

\vspace*{2mm}

\centering
\renewcommand{\arraystretch}{1.8}
\begin{tabular}{|c|}
\hline
Admissible action set for the \textit{t}th state \\ \hline \hline
\cellcolor{gray}
$
\begin{aligned}
\left\{ \bm{a} \in \mathcal{A} \, : \right. \, & \left| \bm{p}^\top_t(x_t) \bm{a}^{(n)} - y^{(n)}_t \right| \leq \epsilon^{(n)} , 
\\
& \left. [\bm{a}^{(n)}]_{1:\rho} = \bm{e}^{(n)}_{t-1} , \, \forall n \in [1,N] \right\}
\end{aligned}
$
\\ \hline
$
\left\{ \bm{a} \in \mathcal{A} \, : \, [\bm{a}^{(n)}]_{1:\rho} = \bm{e}^{(n)}_{t-1} , \, \forall n \in [1,N] \right\}
$  \\ \hline
\end{tabular} 
\renewcommand{\arraystretch}{1}
\label{table:policy_variations}
\end{table}

In this work, we are interested in stationary parametric policies of the form 
\begin{equation} \label{eq:policy}
    \bm{\mu}_{\bm{\theta}}(\bm{s}_t) = \text{arg} \underset{\bm{a}\in\mathcal{A}(\bm{s}_t)}{\text{ min }} \left\{ K(\bm{s}_t,\bm{a}) + J_{\bm{\theta}}(\bm{s}_t,\bm{a};\bm{h}_t)\right\} ,
\end{equation}
where $K$ denotes the same cost used in the objective \eqref{seq:policy_search_objective}, and the map $J_{\bm{\theta}}:\mathcal{S}\times\mathcal{A}\times\mathbb{R}^L\to\mathbb{R}$ is a parametric approximation of the cost-to-go \cite{bertsekas2012dynamic}.
Here, $\bm{h}_t\in\mathbb{R}^L$ represents a latent state, common for all $N$ time series at the \textit{t}th time step, and it plays the role of policy memory \cite{peshkin2001learning,schafer2008reinforcement,zhang2016learning}.
Notice that the choice of $J_{\bm{\theta}}$ determines the architecture of the policy, i.e., the relation between the $P$ parameters within $\bm{\theta}$.
In this work, we adopt three possible policy architectures.

First, we consider a \emph{myopic} policy consisting of a parameterless policy architecture where $J_{\bm{\theta}}(\bm{s}_t,\bm{a};\bm{h}_t) = 0$ for all $t$.
The myopic policy is arguably the simplest architecture-wise but ignores the cost-effect of current actions in future decisions.
Second, we suggest an \emph{RNN-based} policy corresponding to the following parametric cost-to-go approximation
\begin{equation} \label{eq:RNN_cost-to-go}
    J_{\bm{\theta}}(\bm{s}_t,\bm{a};\bm{h}_t) = \sum^N_{n=1} \lambda_n \left\Vert \bm{a}^{(n)} -
    \begin{bmatrix}
        \bm{0}_\rho \\
        \bm{r}^{(n)}_t
    \end{bmatrix}
    \right\Vert^2_2 ,
\end{equation}
where the vectors $\bm{r}^{(n)}_t\in\mathbb{R}^\rho$, for all $n\in\mathbb{N}^{[1,N]}$, and $\bm{h}_t$ represent the output value and latent state of an RNN $R_{\bm{\theta}'}:\mathcal{S}\times\mathbb{R}^L\to\mathbb{R}^{N\rho}\times\mathbb{R}^L$, respectively.
They are obtained through the following relation
\begin{equation} \label{eq:rnn}
    R_{\bm{\theta}'}(\bm{s}_t;\bm{h}_t) = 
    \begin{bmatrix}
        \bm{r}_t \\
        \bm{h}_{t+1}
    \end{bmatrix} ,
\end{equation}
where $\bm{r}_t = [{\bm{r}^{(1)}_t}^\top,\dots,{\bm{r}^{(N)}_t}^\top]^\top$, $\bm{\theta}'\in\mathbb{R}^{P-N}$ and $\bm{\theta}=[\lambda_1,\dots,\lambda_N,{\bm{\theta}'}^\top]^\top\in\mathbb{R}^N_+\times\mathbb{R}^{P-N}$.
The vector $\bm{0}_\rho$ is used in \eqref{eq:RNN_cost-to-go} to disregard the components of every $\bm{a}^{(n)}$ that are already decided by imposing the continuity constraints in \eqref{eq:cc_vf}.
The RNN-based policy promotes actions close, in the sense of a Euclidean distance, to the output of a certain RNN, in this case, $R_{\bm{\theta}'}$.
Thus, an RNN that successfully captures the spatiotemporal dynamics of the process generating the multivariate time series can lead to admissible actions with a reduced-cost effect in future decisions.
Finally, we utilize a \emph{batch} policy, or solution, involving an exact cost-to-go approximation.
The name of the policy comes from the fact that you need the whole batch of streaming observations in advance to construct the exact cost-to-go, i.e., solving \eqref{eq:natural_spline_approximation}.
Thus, in practice, it cannot be used under zero-delay requirements.
The batch solution is parameterless, and by definition, it yields the signal reconstruction with the lowest possible cost.

From the presented policy architectures, notice that we can build different policy variations depending on the choice of the cost $K$ and admissible set $\mathcal{A}(\cdot)$.
In this work, we devise a policy yielding consistent signal reconstructions and present another policy variation designed to reconstruct noisy data smoothly but not consistently.
Both presented policy variations, called from now on consistent and smoothing, respectively, are based on the fact that the roughness, as expressed in \eqref{seq:roughness}, can be equivalently computed from a trajectory of state-action pairs, as we have shown in our previous work (\cite[Proposition 2]{ruizmoreno2023trainable}).
Specifically, the roughness of a spline-based signal estimate $\bm{f}_T = [f^{(1)}_T,\dots,f^{(N)}_T]^\top$ over the domain $(0,X]$ can be computed as
\begin{equation}
    \sum^N_{n=1}\int^X_0\left(D^\rho_x f^{(n)}_T(x)\right)^2 \, dx = \sum^T_{t=1}\sum^N_{n=1} {\bm{a}^{(n)}_t}^\top \bm{M}_t \bm{a}^{(n)}_t ,
\end{equation}
where the elements of $\bm{M}_t\in\bm{S}^{2\rho}_+$ are obtained through 
\begin{equation} \label{eq:definition_M}
\left[\bm{M}_t\right]_{i,j} =
\begin{cases}
    0 & \text{if } i\leq \rho \text{ or } j\leq \rho , \\
    u_t^{i+j-2\rho-1}\frac{\prod^\rho_{k=1}(i-k)(j-k)}{i+j-2\rho -1} & \text{otherwise} ,
\end{cases}
\end{equation}
with $u_t \triangleq x_t - x_{t-1}$, except $u_1 = x_1$.
Similarly, we can construct a roughness-aware penalization from a cost of the form
\begin{equation}
    K(\bm{s}_t,\bm{a}_t) = \sum^N_{n=1} \kappa \left( \bm{s}^{(n)}_t,\bm{a}^{(n)}_t \right) ,
\end{equation}
where $\bm{s}^{(n)}_t$ is an auxiliar per time-series state constructed from the \textit{t}th state, $\bm{s}_t = [x_{t-1},\bm{o}^\top_t,\bm{e}^\top_{t-1}]^\top$, as $\bm{s}^{(n)}_t = [x_{t-1},x_t,y^{(n)}_t,\epsilon^{(n)},{\bm{e}^{(n)}_{t-1}}^\top]^\top$ and where $\kappa$ denotes a per time-series cost.
Table \ref{table:policy_variations} summarizes our proposed policy variations.

\begin{figure}
    \centering
    \includegraphics[width=0.95\columnwidth]{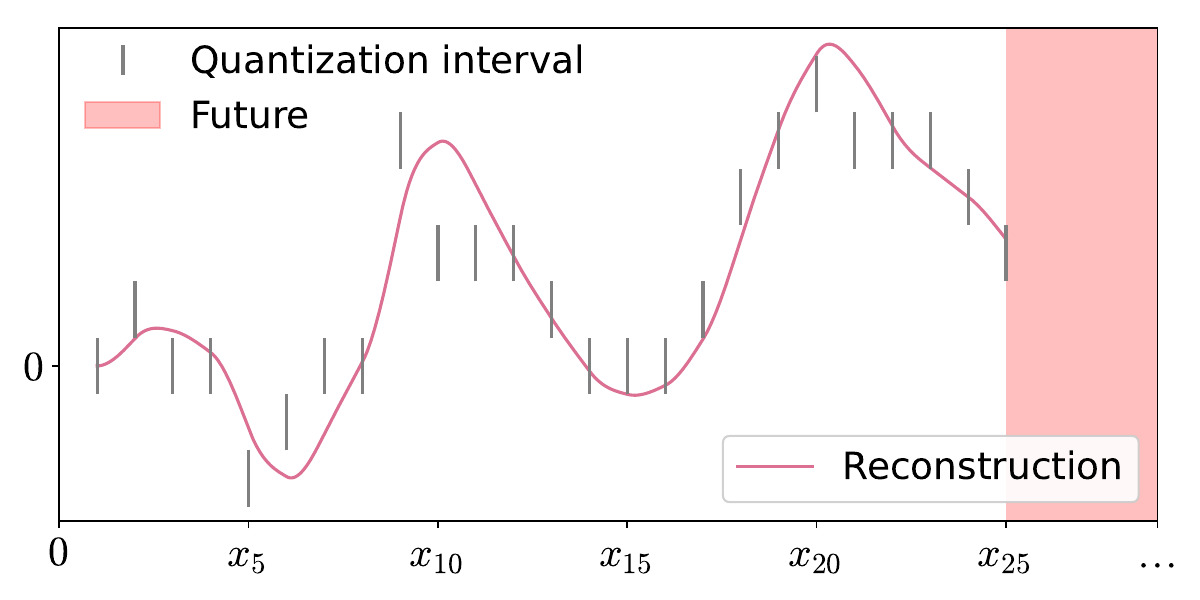}
    \caption{Snapshot of an ongoing signal reconstruction using the smoothing policy variation.}
    \label{fig:smoothing_reconstruction}
\end{figure}

Briefly, the consistent policy variation ensures smoothness while enforcing that the reconstructed signal passes through the quantization intervals, thus yielding a consistent signal reconstruction.
On the other hand, the smoothing policy variation extends the smoothing interpolation approach, introduced in our previous work \cite{ruizmoreno2023trainable}, to multivariate time series.
Specifically, it aims to reduce the squared residuals with the center of the intervals while promoting smoothness, but it does not enforce that the ongoingly reconstructed signal passes through the quantization intervals, as illustrated in Fig. \ref{fig:smoothing_reconstruction}, hence neither consistency.

\subsection{Policy evaluation} \label{ssec:policy_evaluation}
\begin{figure}
    \centering
    \includegraphics[width=0.75\columnwidth]{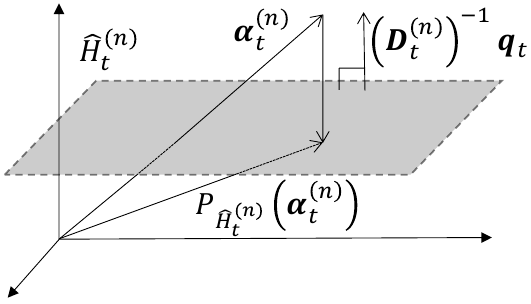}
    \caption{Conceptualization of the hyperplane projection $P_{\hplus}$ in the inner product space $\left( \mathbb{R}^\rho , \langle\cdot , \cdot\rangle_{\bm{D}^{(n)}_t} \right)$.}
    \label{fig:hyperplane_projection}
\end{figure}

\begin{table}
\caption{Terms of the quadratic form for the smoothing policy variation. This notation is shared with the rest of the paper with the incorporation of $\bm{P}_t \triangleq \bm{p}_t(x_t)\bm{p}_t(x_t)^\top$ and $\bm{v}^{(n)}_t \triangleq [\bm{0}^\top_{\rho},{\bm{r}^{(n)}_t}^\top]^\top$.}
\centering
\renewcommand{\arraystretch}{1.7}
\begin{tabular}{|c|c|c|}
\cline{2-3}
\multicolumn{1}{l|}{} & $\bm{A}^{(n)}_t$   & $\bm{b}^{(n)}_t$   \\  \cline{2-3} \noalign{\vskip\doublerulesep \vskip-\arrayrulewidth} \hline
\textbf{Myopic} & $\bm{P}_t + \eta \bm{M}_t$ & $-2y^{(n)}_t\bm{p}_t(x_t)$ \\ \hline 
\textbf{RNN} & $\bm{P}_t + \eta \bm{M}_t + \lambda_n \bm{I}_{2\rho}$ & $-2(y^{(n)}_t\bm{p}_t(x_t) + \lambda_n\bm{v}^{(n)}_t)$ \\ \hline
\end{tabular}
\renewcommand{\arraystretch}{1}
\label{table:smoothing_quadratic}
\end{table}
\begin{table}
\caption{Terms of the quadratic form for the consistent policy variation. Same notation as Table \ref{table:smoothing_quadratic}.}
\centering
\renewcommand{\arraystretch}{1.7}
\begin{tabular}{|c|c|c|}
\cline{2-3}
\multicolumn{1}{l|}{} & $\bm{A}^{(n)}_t$   & $\bm{b}^{(n)}_t$   \\  \cline{2-3} \noalign{\vskip\doublerulesep \vskip-\arrayrulewidth} \hline
\textbf{Myopic} & $\bm{M}_t$ & $\bm{0}_{2\rho}$ \\ \hline 
\textbf{RNN} & $\bm{M}_t + \lambda_n \bm{I}_{2\rho}$ & $-2\lambda_n\bm{v}^{(n)}_t$ \\ \hline
\end{tabular}
\renewcommand{\arraystretch}{1}
\label{table:consistent_quadratic}
\end{table}

\begin{figure*}
    \centering
    \includegraphics[width=0.95\linewidth]{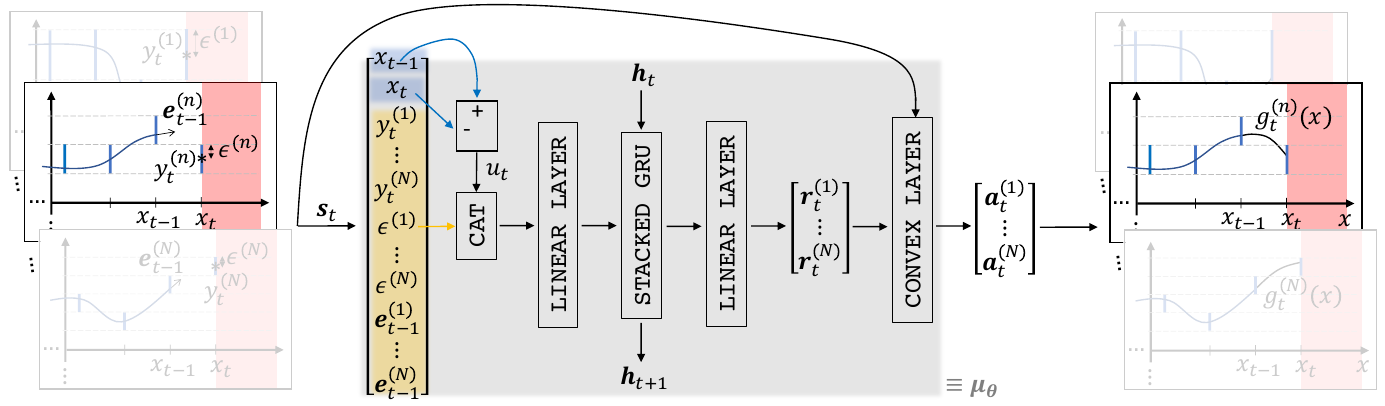}
    \caption{Representation of a \textit{t}th forward pass of our RNN-based policy. CAT refers to a vector concatenation layer. Evaluating the policy consists of solving a convex optimization problem with respect to $\bm{a}$, see \eqref{eq:policy} and Sec. \ref{ssec:policy_evaluation}. Therefore, its last step can be seen as a convex optimization layer \cite{agrawal2019differentiable}.}
    \label{fig:rnn}
\end{figure*}

By design, evaluating the considered policies involves solving optimization problem \eqref{eq:policy}.
The myopic and RNN-based policy architectures, in any of their variations given in Table \ref{table:policy_variations}, can be equivalently evaluated by solving the following quadratic convex problem
\begin{equation} \label{eq:policy_quadratic}
    \bm{\mu}_{\bm{\theta}}(\bm{s}_t) = \text{arg} \underset{\bm{a}\in\mathcal{A}(\bm{s}_t)}{\text{min}} \left\{ \sum^N_{n=1} {\bm{a}^{(n)}}^\top \bm{A}^{(n)}_t \bm{a}^{(n)} + {\bm{b}^{(n)}_t}^\top\bm{a}^{(n)} \right\} ,
\end{equation}
where the terms $\bm{A}^{(n)}_t \in \bm{S}_+^{2\rho}$ and $\bm{b}^{(n)}_t \in \mathbb{R}^{2\rho}$ depend on the chosen policy architecture and variation, as displayed in Tables \ref{table:smoothing_quadratic} and \ref{table:consistent_quadratic}.
The policy evaluation in (\ref{eq:policy_quadratic}) is presented as an  intermediate step for clarity.
Then, after some algebraic steps, it can be equivalently reformulated as
\begin{equation} \label{eq:policy_distance}
    \bm{\mu}_{\bm{\theta}}(\bm{s}_t) = \text{arg} \underset{\bm{a}\in\mathcal{A}(\bm{s}_t)}{\text{min}} \left\{ \sum^N_{n=1} \Vert \bm{a}^{(n)} - \bm{\alpha}^{(n)}_t \Vert^2_{\bm{D}^{(n)}_t} \right\}
\end{equation}
where every $\bm{D}^{(n)}_t \in \bm{S}^{\rho}_{++}$ and $\bm{\alpha}^{(n)}_t \in \mathbb{R}^{\rho}$ are computed from the elements of the \textit{t}th state, through $\bm{A}^{(n)}_t$ and $\bm{b}^{(n)}_t$, as
\begin{equation}
    \bm{D}^{(n)}_t = \bm{B}_2^\top \bm{A}^{(n)}_t \bm{B}_2 ,
\end{equation}
and
\begin{equation}
    \bm{\alpha}^{(n)}_t = - {\bm{D}^{(n)}_t}^{-1} \left( \bm{B}_2\bm{A}^{(n)}_t\bm{B}_1\bm{e}^{(n)}_{t-1} + \frac{1}{2}\bm{B}_2^\top\bm{b}^{(n)}_t \right) ,
\end{equation}
for all $n\in\mathbb{N}^{[1,N]}$ and $t\in\mathbb{N}^{[1,T]}$.
The auxiliary matrices $\bm{B}_1 \triangleq [\bm{I}_{\rho}, \bm{0}_{\rho\times\rho}]^\top \in \mathbb{R}^{2\rho\times\rho}$ and $\bm{B}_2 \triangleq [\bm{0}_{\rho\times\rho},\bm{I}_{\rho}]^\top \in \mathbb{R}^{2\rho\times\rho}$ are defined for notational brevity.
The evaluation form in (\ref{eq:policy_distance}) is analytically preferable since it corresponds to a minimum norm problem or a hyperslab projection problem \cite{slavakis2014online,ruiz2021tracking}, for the smoothing or consistent policy variations, respectively.
This is because the aforementioned problems admit the following closed-form solution
\begin{equation}
    \bm{\mu}_{\theta}(\bm{s}_t) = 
    \begin{bmatrix}
        {\bm{a}^{(1)}_t}^* \\
        \vdots \\
        {\bm{a}^{(N)}_t}^* 
    \end{bmatrix} ,
    \text{ with } \,
    {\bm{a}^{(n)}_t}^* =
    \begin{bmatrix}
        \bm{e}^{(n)}_{t-1} \\
        \zeta^{(n)}_t\left(\bm{\alpha}^{(n)}_t\right)
    \end{bmatrix} ,\, \forall n,
\end{equation}
where $\zeta^{(n)}_t:\mathbb{R}^\rho\to\mathbb{R}^\rho$ is the identity operator for the smoothing policy variation and the projection onto the hyperslab
\begin{equation}
    \hslab \triangleq \left\{ \bm{\beta} \in \mathbb{R}^\rho : \left| \bm{\beta}^\top\bm{q}_t - w^{(n)}_t \right| \leq \epsilon^{(n)} \right\},
\end{equation}
for the consistent one, where $\bm{q}_t\triangleq [\bm{p}_t(x_t)]_{\rho+1:2\rho}$ and $w^{(n)}_t \triangleq y^{(n)}_t - {\bm{e}^{(n)}_{t-1}}^\top [\bm{p}_t(x_t)]_{1:\rho}$ for all $t\in\mathbb{N}^{[1,T]}$ and $n\in\mathbb{N}^{[1,N]}$.
Every hyperslab $\hslab$ can be visualized as the set of all points which belong between and onto the hyperplanes
\begin{subequations}
\begin{align}
    \hplus \triangleq \left\{ \bm{\beta}\in\mathbb{R}^\rho : \bm{\beta}^\top\bm{q}_t = w^{(n)} + \epsilon^{(n)} \right\} , \\
    \hmin \triangleq \left\{ \bm{\beta}\in\mathbb{R}^\rho : \bm{\beta}^\top\bm{q}_t = w^{(n)} - \epsilon^{(n)} \right\} .
\end{align}
\end{subequations}

Based on this observation, the projection onto the hyperslab $\hslab$ can be computed as 
\begin{equation} 
\zeta^{(n)}_t(\bm{\beta}) =
\begin{cases}
    P_{\hplus}(\bm{\beta}), & \text{if } \bm{\beta}^\top\bm{q}_t > w^{(n)}_t + \epsilon^{(n)},  \\
    \bm{\beta}, & \text{if } \left| \bm{\beta}^\top\bm{q}_t - w^{(n)}_t \right| \leq \epsilon^{(n)}, \\
    P_{\hmin}(\bm{\beta}), & \text{if } \bm{\beta}^\top\bm{q}_t < w^{(n)}_t - \epsilon^{(n)} ,
\end{cases}
\end{equation}
where $P_{\hplus}:\mathbb{R}^\rho\to\hplus$ and $P_{\hmin}:\mathbb{R}^\rho\to\hmin$ denote the projections onto the hyperplanes $\hplus$ and $\hmin$, respectively.
These projections can be computed in closed form as
\begin{subequations} 
\begin{align}
    P_{\hplus}(\bm{\beta}) = \bm{\beta} - \left( \bm{\beta}^\top\bm{q}_t - w^{(n)}_t - \epsilon^{(n)} \right) \frac{ \left(\bm{D}^{(n)}_t\right)^{-1} \bm{q}_t }{\bm{q}_t^\top \left(\bm{D}^{(n)}_t\right)^{-1} \bm{q}_t} , \label{seq:hyperplane+_projection} \\
    P_{\hmin}(\bm{\beta}) = \bm{\beta} - \left( \bm{\beta}^\top\bm{q}_t - w^{(n)}_t + \epsilon^{(n)} \right) \frac{ \left(\bm{D}^{(n)}_t\right)^{-1} \bm{q}_t }{\bm{q}_t^\top \left(\bm{D}^{(n)}_t\right)^{-1} \bm{q}_t} .
\end{align}
\end{subequations}
See Fig. \ref{fig:hyperplane_projection} for some intuition behind the closed-form projection in \eqref{seq:hyperplane+_projection}. 
The same reasoning applies to the other hyperplane. 

\subsection{Policy training}
Unlike the myopic policy architecture and batch solution, which are parameterless, the RNN-based architecture requires to be tuned, in this case, through \eqref{eq:policy_training}, for adequate performance.
Since evaluating the RNN-based policy can be seen as performing a forward pass from a deep learning point of view \cite{goodfellow2016deep}, the policy training problem \eqref{eq:policy_training} can be tackled via backpropagation through time \cite{werbos1990backpropagation}.
Here, the closed-form evaluation described in Sec. \ref{ssec:policy_evaluation} allows us to compute and propagate the gradient of the \textit{t}th action $\bm{a}_t$ with respect to the parameters contained in $\bm{\theta}$ without the need of unrolling numerical optimizers \cite{monga2021algorithm} or other specialized tools \cite{agrawal2019differentiable}.
An overall scheme of our chosen RNN-based policy and its components can be found in Fig. \ref{fig:rnn}.

\section{Error-rate decay behavior} \label{sec:empirical}
\begin{figure}
    \centering
    \includegraphics[width=0.65\columnwidth]{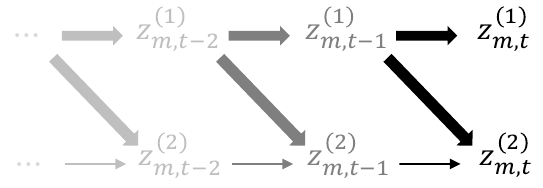}
    \caption{Scheme of the acyclic VAR(1) process generating the $2$ series of knots for any \textit{m}th sequence. The thickness of the arrows represents the magnitude of the autoregressive parameters.}
    \label{fig:VAR_scheme}
\end{figure}

\begin{figure}
    \centering
    \includegraphics[width=0.9\columnwidth]{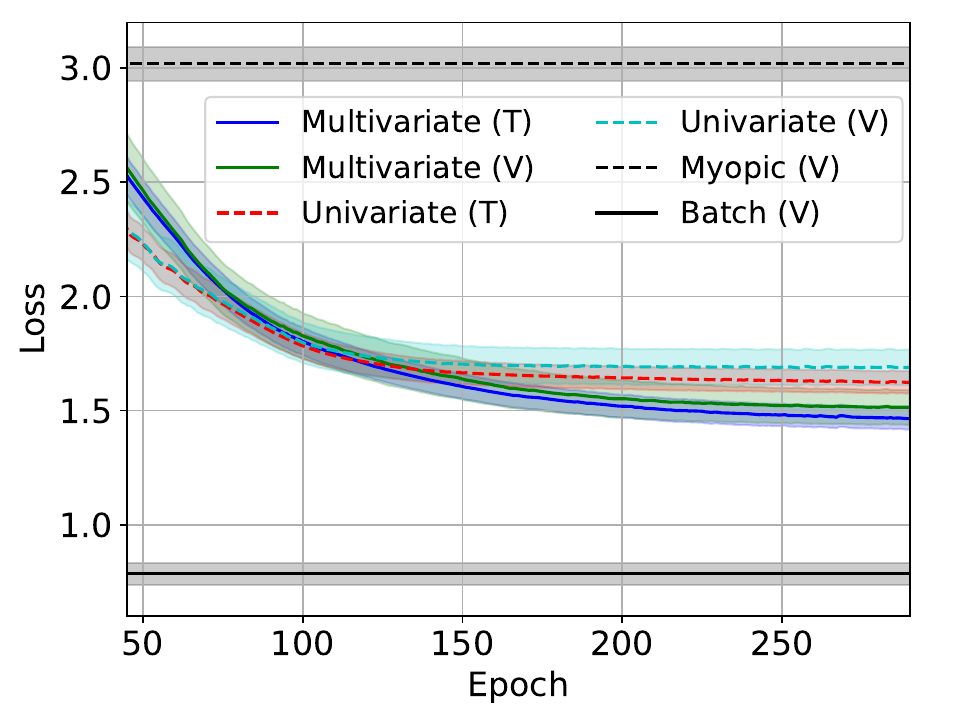}
    \caption{Training-validation curves for the consistent variation of the RNN-based policy. The loss metric is the average total cost per function section computed from the per time-series cost presented in Table \ref{table:policy_variations}.}
    \label{fig:training-validation_curves}
\end{figure}

\begin{figure*}
\begin{subfigure}{0.66\textwidth}
    \centering
    \includegraphics[width=0.49\linewidth]{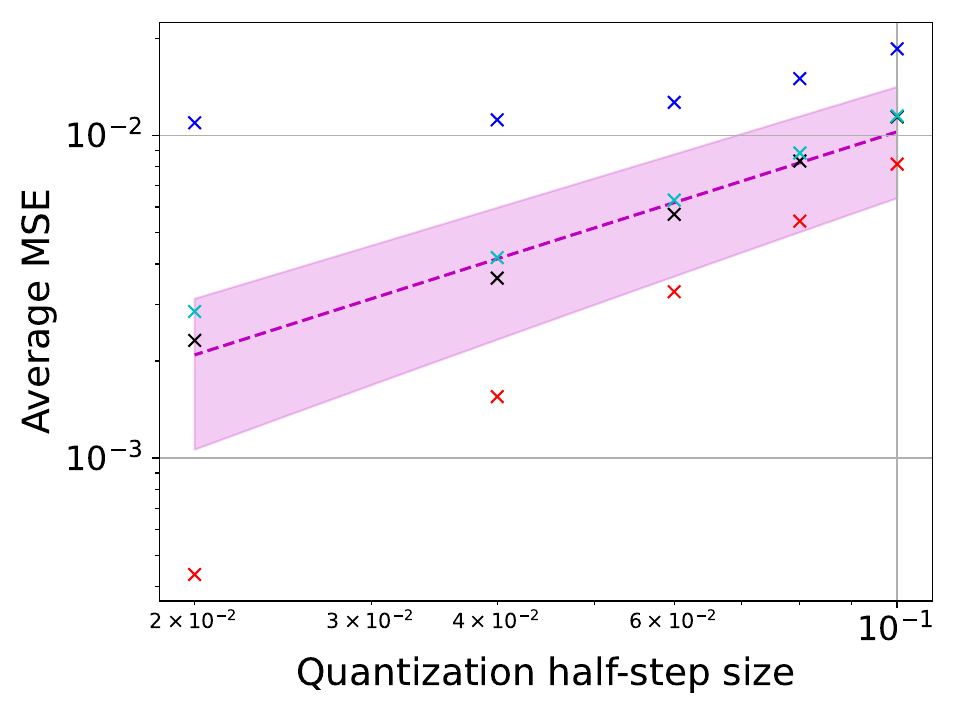}
    \includegraphics[width=0.49\linewidth]{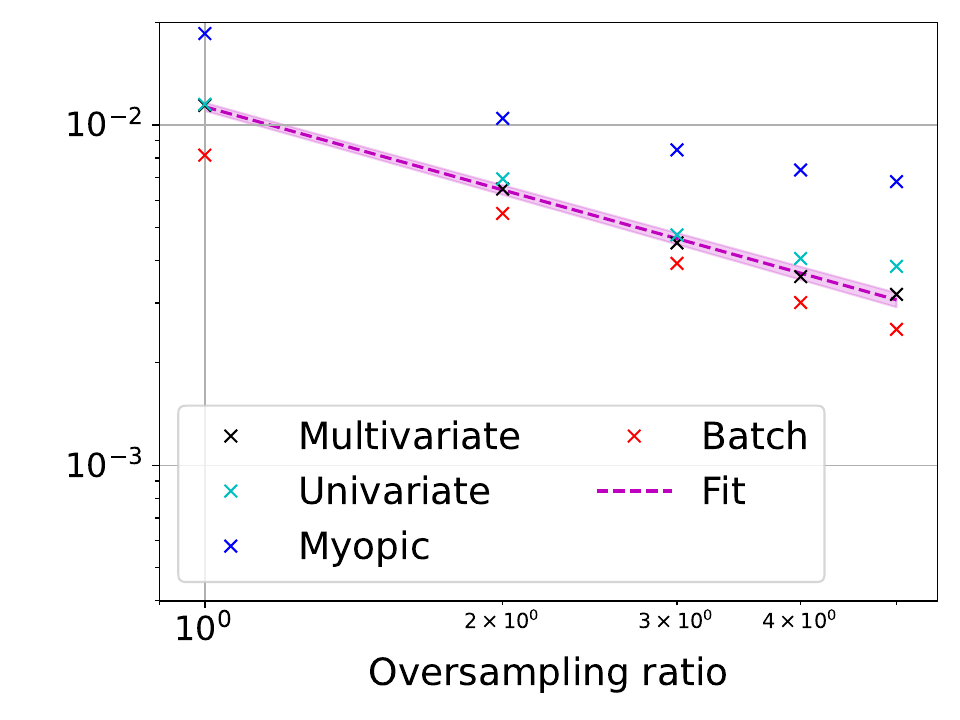}
    \caption{Consistent policy variation.}
    \label{subfig:decay_consistent}
\end{subfigure}
\begin{subfigure}{0.33\textwidth}
    \centering
    \includegraphics[width=0.98\linewidth]{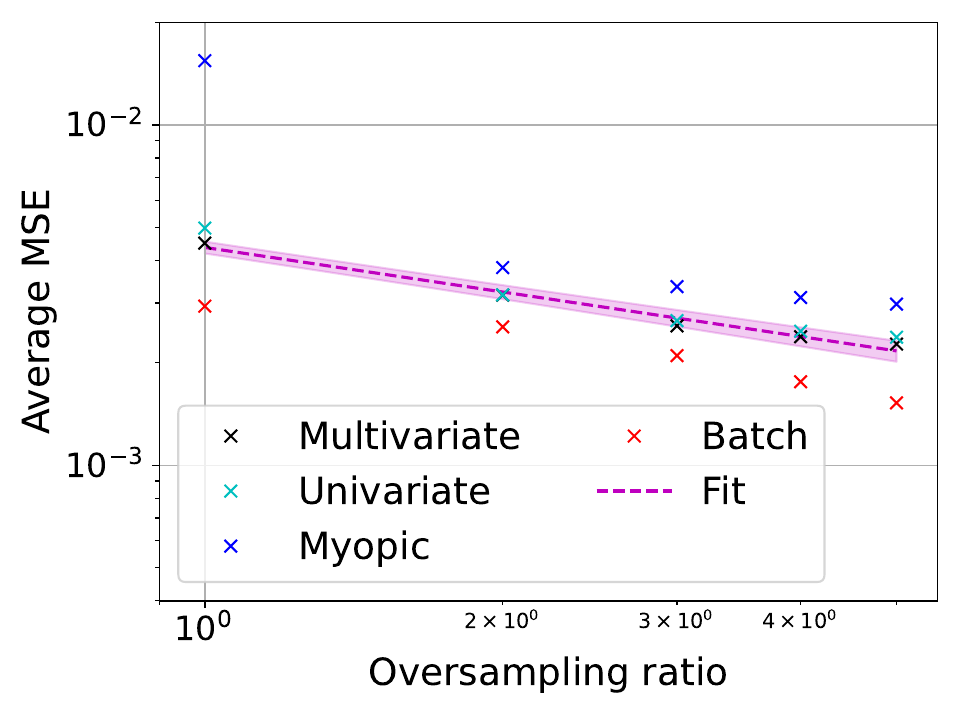}
    \caption{Smoothing policy variation.}
    \label{subfig:decay_smoothing}
\end{subfigure}
\caption{Average MSE decay curves over the test set in log-log scale. The fitting curves are log-linear functions adjusted from the multivariate average MSE values. The shaded area represents the confidence in the slope of the fit. The values in Fig. \ref{subfig:decay_smoothing} have been computed for an $\eta=0.001$.}
\label{fig:decay}
\end{figure*}

In this section, we experimentally analyze the signal reconstruction error-rate decay incurred by the consistent variation of the RNN-based policy (see Sec. \ref{ssec:architecture_and_variation}), with respect to the oversampling ratio and quantization step size of the acquisition system.
To this end, we first set a specific function space of multivariate smooth signals, as described in Sec. \ref{ssec:signal_space}.
For simplicity, we fix the dimensionality of the multivariate signal to $N=2$, and specify the roughness by choosing $\rho=2$.
Then, we describe a form to generate multivariate signals from $\mathcal{V}^2_2$.
These multivariate signals, along with an acquisition system with given oversampling ratio and quantization stepsize values, shape the time series data of the experiment.
Separately, we use the myopic policy architecture as a benchmark, or base method, which does not exploit the spatiotemporal dependencies among time series, and the batch solution as a baseline, or best possible performance indicator.
The central goal of this section is to compare the performance of the RNN-based policy architecture that considers the possible spatial relations between the time series (multivariate) against a policy based on the same RNN architecture but ignoring the spatial relations (univariate). This is done for both policy variations: consistent and smoothing.

\subsection{Data generation}
Unlike real-world acquired signals, which usually include uncertainty due to unidentified external noise sources or lack of knowledge about their underlying generating process, synthetic signals can be generated in a controlled environment.
On the one hand, synthetic data is usually a simplified version of real-world data, hence not always accurately representing the real world.
On the other hand, the mechanisms used to generate synthetic data are fully known.
This is especially convenient in studies involving irreversible transformations such as quantization since there is no other unidentified source of error than the quantization and reconstruction methods at work.
In this paper, we use synthetically generated multivariate time series to delimit, as much as possible, the dependencies between the reconstruction error and the reconstruction method used.
Moreover, the synthetic data we use is complex enough to accommodate the smooth behavior as well as the spatiotemporal dependencies assumed in Sec. \ref{ssec:signal_space}.

First, we generate $M=288$ sequences composed of $N=2$ series of $T_*=100$ knots each, where every knot has been computed recurrently from a stable
vector autoregressive VAR(1) process $\bm{z}_{m,t} = \bm{\Phi}\bm{z}_{m,t-1} + \textbf{w}_{m,t}$ with autoregressive parameters $\bm{\Phi}\in\mathbb{R}^2$, being $[\bm{\Phi}]_{1,2}=0$ to avoid self feedback, Gaussian innovation $\textbf{w}_{m,t} \sim \mathcal{N}\left(\bm{0},\text{diag}([0.1,0.1]^\top)\right)$ and where $\bm{z}_{m,0} = \bm{0}_2$ for all $m\in\mathbb{N}^{[1,M]}$.
In this way, the resulting sequences of knots inherit the spatiotemporal relations of the VAR(1) generator process, as illustrated in Fig. \ref{fig:VAR_scheme}.
Second, each series of knots is arranged uniformly, with a unitary period $1/\nu^* = 1$, and is interpolated with cubic natural splines, which are the smoothest interpolators in terms of the $\rho=2$ roughness \cite{wood2006generalized}.
Lastly, the resulting continuous signals are uniformly sampled at a frequency $\nu=R\nu^*=R$ with a natural oversampling ratio $R$ spanning from $1$ to $5$ and quantized by a uniform midtread quantizer, with a half-step size $\epsilon\in\{0.1,0.08,0.06,0.04,0.02\}$.
As a result, we generate a dataset consisting of $288$ multivariate time series 
with elements of the form $\tilde{y}^{(n)}_{m,t} = \text{round}(z^{(n)}_{m,t}/2\epsilon)2\epsilon$ for every $n\in\{1,2\}$, $m\in\mathbb{N}^{[1,288]}$ and $t\in\mathbb{N}^{[1,T]}$, where $T=100R$.

\subsection{Experimental setup}
The dataset is divided into $196$ samples for training, $64$ for validation, and $32$ for testing.
Then, to avoid data leaking, the whole dataset is standardized along each time series with the mean $\gamma_\text{tr}^{(n)}$ and standard deviation $\sigma_\text{tr}^{(n)}$ of the training partition.
Explicitly, $y^{(n)}_{m,t} = (\Tilde{y}^{(n)}_{m,t} - \gamma_\text{tr}^{(n)})/\sigma_\text{tr}^{(n)}$ and $\epsilon^{(n)} = \epsilon/\sigma_\text{tr}^{(n)}$ for all $n\in\{1,2\}$, $m\in\mathbb{N}^{[1,288]}$ and $t\in\mathbb{N}^{[1,100R]}$, resulting in quantization intervals as described in Sec. \ref{ssec:acquisition_system}.
The RNN-based policy architecture consists of $2$ stacked gated recurrent units (GRU) layers \cite{cho2014properties}, with a latent state of size $L=48$ and an input of size $32$. 
The input and output layers are set as linear layers matching the required dimensionality, see Fig. \ref{fig:rnn} for a visual representation, which accordingly varies depending on whether we reconstruct multivariate time series as a multivariate signal or as separate individual univariate signals.
Lastly, the RNN-based policies have been trained using the adaptive moments (Adam) optimizer \cite{kingma2014adam}, with an exponential decay rate for the first moment estimates of $\beta_1 = 0.9$ and the second-moment estimates of $\beta_2 = 0.999$, without weight decay, a learning rate of $0.002$ and with a gradient norm clip value of $0.1$ over mini-batches of $32$ samples.

\subsection{Results and discussion}
The training-validation curves obtained for the consistent variation of the RNN-based policy, from data acquired with $R=1$ and $\epsilon=0.1$, are presented in Fig. \ref{fig:training-validation_curves}.
As expected, the consistent RNN-based policy outperforms the consistent myopic policy for both the multivariate and univariate reconstruction cases.
Moreover, the multivariate reconstruction outperforms the univariate one, illustrating the model capabilities to successfully exploit spatial dependencies (and not only the temporal ones) from multivariate time series.

Regarding the reconstruction error-rate decay behavior, we have computed the average MSE over all time series in the test set.
This has been done for every considered policy architecture (once they are trained, if applicable) and policy variation.
As shown in Fig. \ref{fig:decay}, we have presented the results by grouping the consistent and smoothing policy variations versus the quantization half-step size $\epsilon$ (with fixed $R=1$) or versus the oversampling ratio $R$ (with fixed $\epsilon=0.1$).
The slope of the fitting curve reflects the average error-rate decay in logarithmic scale incurred by the multivariate reconstructions of the RNN-based policies. 
In the case of the consistent policy variation, the module of the quotient between the slope obtained by varying the quantization step size ($0.99\pm0.09$) over the slope obtained by increasing the oversampling ratio ($-0.812\pm0.008$) is $1.22 \pm 0.11$.
This quotient ratio is only (at most within) two standard deviations from a unit value, suggesting that the balanced error-rate decay property can be achieved for consistent reconstructions even under zero-delay requirements.
As to the smoothing policy variation (with slope $-0.435\pm0.002$), we are interested in its relative performance with respect to the consistent one. 
Here, the quotient ratio between the slopes of the fits is $1.87 \pm 0.02$, meaning that the error-rate decay for the consistent policy variation is nearly doubled with respect to the smoothing one.
Again as expected, a property of offline consistency manifests even under zero-delay requirements.

\section{Conclusion} \label{sec:conclusion}
This paper designs a method for zero-delay signal reconstruction from streaming multivariate time series of quantization intervals, both for the smoothing and consistent discussed variants.
The error-rate decay behavior of the proposed methods has been empirically analyzed showing that the error incurred by the consistent variant decreases nearly twice (in logarithmic scale) as fast as the error incurred by the (non-consistent) smoothing variant, as the oversampling ratio is increased.
Moreover, the former error rate is practically the same as the rate the error decreases when the quantization step size is reduced, which is advantageous from an implementation perspective.
Finally, we also observe that successfully exploiting the spatiotemporal dependencies of the streaming quantized data, in this case via policy training, helps to reduce the average signal reconstruction error.

    \bibliography{references}

\end{document}